\pdfoutput=1
\documentclass[a4paper]{article}

\usepackage{INTERSPEECH2021}

\usepackage{times}
\usepackage{latexsym}
\usepackage[T1]{fontenc}
\usepackage[utf8]{inputenc}

\usepackage{multirow}
\usepackage{booktabs}
\usepackage{hyperref}
\usepackage{todonotes}
\usepackage{xspace}
\usepackage{graphicx}
\usepackage{caption}
\usepackage{subcaption}
\usepackage{paralist}
\usepackage{amsmath}
\usepackage{amssymb}
\usepackage{stmaryrd}
\usepackage{float}

\newcommand{\rnr}{{\textsc{RnR}}\xspace}
\newcommand{\redalph}{{$\Re$}\xspace}
\newcommand{\rone}{{$\rho_1$}\xspace}
\newcommand{\rtwo}{{$\rho_2$}\xspace}

\title{Reduce and Reconstruct: ASR for Low-Resource Phonetic Languages}

\name{Anuj Diwan, Preethi Jyothi}
\address{
Department of Computer Science and Engineering, Indian Institute of Technology Bombay, India}
\email{\{anujdiwan,pjyothi\}@cse.iitb.ac.in}

\begin{document}

\maketitle
\begin{abstract}
  %End-to-end automatic speech recognition (ASR) systems are data intensive and hence underperform in low-resource settings. 
 This work presents a seemingly simple but effective technique to improve low-resource ASR systems for phonetic languages. By identifying sets of acoustically similar graphemes in these languages, we first reduce the output alphabet of the ASR system using linguistically meaningful reductions and then reconstruct the original alphabet using a standalone module. We demonstrate that this lessens the burden and improves the performance of low-resource end-to-end ASR systems (because only reduced-alphabet predictions are needed) and that it is possible to design a very simple but effective reconstruction module that recovers sequences in the original alphabet from sequences in the reduced alphabet. We present a finite state transducer-based reconstruction module that operates on the 1-best ASR hypothesis in the reduced alphabet. We demonstrate the efficacy of our proposed technique using ASR systems for two Indian languages, Gujarati and Telugu. With access to only 10 hrs of speech data, we obtain relative WER reductions of up to $7\%$ compared to systems that do not use any reduction.  %Rest will be filled in after the latest numbers come out.
  \end{abstract}

% \iffalse
\section{Introduction}

End-to-end (E2E) automatic speech recognition (ASR) systems are becoming an
increasingly popular choice for ASR modeling~\cite{chiu2018state}. E2E systems
directly map speech to sequences of graphemes or subword units derived from
graphemes. %thus allowing for pronunciation lexicons to be bypassed. 
This direct treatment of the ASR problem makes E2E modeling an attractive choice for low-resource and high-resource languages alike. 
%for which it may be hard to acquire resources like pronunciation dictionaries that require linguistic expertise. 
However, E2E ASR systems are very data intensive and consequently tend to underperform on low-resource languages for which labelled speech data is scarce. Identifying techniques that can help boost E2E performance for low-resource languages is of great interest. We offer one such approach that we refer to as \emph{Reduce and Reconstruct} (\rnr). 

To motivate \rnr, we start with reiterating the goal of E2E models. These models aim at learning direct mappings from speech to graphemes. In low-resource settings, there may not be sufficient amounts of data to learn these mappings reliably. This issue could be alleviated by meaningfully reducing the size of the output vocabulary using a simple rule-based system that is linguistically motivated. This is the \textit{Reduce} step in \rnr. The resulting alphabet should be compact, while also being sufficiently discriminative across speech sounds. With this reduced alphabet in place, we now have an easier task to be learned by the E2E models. Since the predictions from the E2E models will be in the reduced alphabet, one would require an additional reconstruction module (i.e. the \textit{Reconstruct} step in \rnr) to recover the original grapheme sequence. We tackle the reconstruction problem using WFSTs, since the \textit{Reduce} step is a deterministic mapping from the original grapheme alphabet to the reduced alphabet which can be appropriately modelled using WFSTs. We note here that \rnr is well-suited for E2E models that learn direct mappings from speech to graphemes. For traditional cascaded ASR systems, a reduced alphabet might actually be counterproductive with introducing more confusability across words in the lexicon.

Consider the following examples that illustrate the utility of \rnr. Suppose our reference text is ``call the bus" and we train an ASR model with a reduced output vocabulary \redalph that uses a single character $\zeta$ to represent both ``c" and ``k"
in the original English alphabet. If the ASR model is able to accurately
predict ``$\zeta$all", then reconstruction will trivially map ``$\zeta$all" to
``call" since ``kall" is not a valid word in English. A more interesting case
is when there is more than one valid word that a reduced word form can map to.
Say our reference text is ``game is on" and \redalph has a single character
$\kappa$ that maps to ``c", ``k" and ``g".  If the ASR model correctly predicts
$\kappa$, ``$\kappa$ame" could be mapped to either ``came" or ``game". However,
the language model in the reconstruction module should accurately pick ``game
is on" as the more likely prediction given its higher likelihood.

In this work, we consider two low-resource Indian languages, Gujarati and Telugu, to demonstrate the use of \rnr. \rnr can be easily applied to any language (like Telugu, Gujarati, etc.) that is largely phonetic with one-to-one mappings between phones and graphemes. We first identify sets of acoustically similar graphemes in each language to produce an appropriate reduction, train an ASR system with the reduced vocabulary and finally reconstruct the best ASR prediction using an FST-based reconstruction module. 

% The idea of \rnr is even more attractive in the context of Indian
% languages that are highly phonetic in nature.% and have phonologies that are structured in similar ways. 
% %For example, aspirated variants of both voiced and voiceless consonants appear in many Indian languages. 
% This allows us to easily come up
% with an appropriate reduction for each language which we will discuss in more
% detail in Section~\ref{sec:ourapproach}. Our approach is particularly attractive in that it only requires a one-time effort of identifying sets of acoustically similar graphemes in a given language to produce an appropriate reduction.

% In summary, our overall contributions are:
% \begin{itemize}
% \item We propose a new paradigm reduce and reconstruct, \rnr, to be used with E2E ASR in low-resource settings.
% \item We present a finite state transducer (FST)-based reconstruction technique.
% \item We show the benefits of \rnr with E2E ASR systems for two low-resource languages, Telugu and Gujarati.
% \end{itemize}

\section{Related Work} 
Our line of work is closely related to error correction models for ASR that explicitly correct errors made by the ASR system~\cite{ringger1996error}. Several such efforts for error correction of ASR systems have been carried out in prior work; \cite{Errattahi2018} presents a review. Past work has looked at reordering ASR hypotheses using machine translation-inspired techniques~\cite{cucu2013statistical,d2016automatic},  leveraging contextual information using recurrent models and rescoring confusion networks generated by the ASR system~\cite{sarma2004context,jung2004speech,tam2014asr,nakatani2013two,byambakhishig2014error,fusayasu2015word}. Our proposed framework is different in that it is not strictly a postprocessing technique and modifies the ASR system itself to use a reduced alphabet. More recent work has looked at correcting errors made by E2E models by training either sequence-to-sequence or transformer-based models~\cite{Guo2019,9053051,zhang2019automatic}.

Another different but related line of work investigates the ways in which sounds from different languages can be merged effectively when building multilingual ASR systems~\cite{melnar2003phone}. Attempts at redefining the phone set in a data-driven manner have also been made for monolingual ASR systems~\cite{singh2000structured}. Recent work has looked at the influence of merging phones for code-switched speech recognition~\cite{sivasankaran-etal-2018-phone} and building language-agnostic multilingual systems by mapping graphemes in Indian languages to a single script~\cite{9053443}.

\section{\rnr: Reduce and Reconstruct}
\label{sec:ourapproach}

%\rnr proceeds through the following steps. 
We first devise a many-to-one reduction mapping where each grapheme in the original alphabet is mapped to a reduced grapheme, thus creating a new reduced alphabet \redalph. An ASR model is trained using transcriptions in \redalph. The 1-best decoded output from this model is fed as input to a reconstruction module that is trained to recover transcriptions in the original alphabet from the reduced alphabet. We note that using the $n$-best decoded outputs ($n = 100$) from the ASR model, for reconstruction, resulted in exactly the same performance as just using the $1$-best, so we use the latter in all our experiments.
%We first describe our choice of reduction functions, followed by the FST-based reconstruction module that we designed.

\subsection{Reduction Functions}

We aim to group together graphemes that are acoustically similar and replace each such set with a reduced grapheme. We manually design such reduction functions  by referring to the phonology of Gujarati and Telugu. This task is simple for phonetic languages, since they have mostly one-to-one mappings between graphemes and phonemes.%
\footnote{Reduction mappings could also be data-driven and learned automatically from speech, which we leave for future work.}
\footnote{In this work, while we evaluate on two languages with phonetic orthographies, it should be possible to use \rnr with any language for which there is sufficient linguistic expertise to deterministically map  its character inventory down to a reduced ``pseudo-phone" set.}
%

%For example, Gujarati has both a voiceless velar stop and an aspirated voiceless velar stop that are acoustically similar and appear as two distinct graphemes. One could aim to learn such mappings automatically from speech data. Given that we are in the low-resource setting where availability of such data is scarce, in this work, we refer to the phonology of Gujarati and Telugu and design reduction functions that are linguistically motivated. This task is also made simpler by the fact that both Gujarati and Telugu are phonetic languages with mostly one-to-one mappings between graphemes and phonemes. For a new language, an appropriate reduction can be formulated by identifying such sets of acoustically similar but linguistically distinct graphemes.

Both Gujarati and Telugu have five different types of plosives that vary in
their place of articulation (labial, alveolar, retroflex, palatal and velar). Each of these plosives can be voiced or unvoiced and further, aspirated or unaspirated.
%Each of these plosives have both voiced and unvoiced variants, and each of these can further be either aspirated or unaspirated. 
For a given place of
articulation, we merge the graphemes corresponding to all four plosives into a
single grapheme in \redalph. %That is, we map each of these four graphemes to a single grapheme in the reduced alphabet \redalph. 
Overall, the graphemes corresponding
to the above-mentioned twenty plosives are reduced to five graphemes in \redalph. There are five graphemes corresponding to nasal sounds in Gujarati and Telugu, which we reduce down to a single grapheme. %In Gujarati, ['સ','શ','ષ'], ['લ','ળ'], and ['ય','હ'] groups are merged as well. Their Telugu counterparts ['స','శ','ష'], ['ల','ళ'], ['య','హ']are also merged. 
In both languages, long and short vowel variants corresponding to a particular sound are merged into a single grapheme. All other graphemes are left as-is in the reduced alphabet.  This reduction will henceforth be referred to as \rone. This reduces the grapheme alphabet size to $27$ from $63$ for Gujarati and to $27$ from $70$ for Telugu.

\subsection{FST-based Reconstruction}

Our reconstruction model is implemented using a cascade of FSTs. This structure is somewhat similar to the decoder in~\cite{7404790}, although the specific FSTs used in our work and the motivation for their design are very different.   
% Figure~\ref{fig:fstcascade} illustrates the FST-based reconstruction pipeline.
The input to the reconstruction model is an ASR hypothesis consisting of a sequence of reduced graphemes. Let this input be represented as a linear chain acceptor $H$. $H$ is transduced to an output FST $O$ using a series of FST compositions, followed by an invocation of the $\mathrm{shortestpath}$ algorithm on the composed FST:
\begin{align*}
O = \mathrm{shortestpath} (H \circ S \circ E \circ L \circ G)
\label{eq:fsts}
\end{align*}
The FSTs $S$, $E$, $L$ and $G$ (described below) are weighted using negative log probability costs, thus making $\mathrm{shortestpath}$ appropriate. The output labels of $O$ will correspond to the best reconstructed word sequence in the original alphabet.\\
%The following sequence of compositions, followed by an invocation of the $\mathrm{shortestpath}$ algorithm on the composed FST, will yield an output FST $O$. The output labels from this shortest path FST will correspond to the reconstructed grapheme sequence in the original alphabet:
%

%\noindent Each FST in Equation~\ref{eq:fsts} is detailed below.
\noindent \textbf{$S$: Reduction FST.} $S$ is a 1-state machine that takes 
reduced graphemes as input and produces original graphemes as its output.
$S$ contains zero-cost transitions corresponding to pairs $(g,\hat{g})$
where $g$ is a reduced grapheme in \redalph and $\hat{g}$ is its counterpart 
in the original alphabet. $H \circ S$ would exhaustively
produce all possible reconstructions of the reduced word forms in $H$.
\vspace{0.3em}

\noindent \textbf{$E$: Edit distance FST.} By examining the ASR errors, we observe that many predicted words differ from the correct words
in only a small number of graphemes. To accommodate this, we design an edit
distance FST $E$ that takes a space-separated grapheme sequence as input and produces as output
all reduced word forms that are within an edit distance of $d$ from each word in the input. The allowable edits are substitutions, insertions and deletions. Each
edit incurs an additive cost $\lambda$. Both $d$ and $\lambda$ are tunable
hyperparameters. $H \circ S \circ E$ will now produce all possible
reconstructions of the reduced word forms in $H$ with zero cost and also other
words with cost $\lambda e$ ($1
\le e \le d$) that are at an edit distance $e$ from the exact reconstructions.
\vspace{0.3em}
    
\noindent \textbf{$L$: Dictionary FST.} The dictionary FST maps sequences of
graphemes to sequences of words. This machine is similar in structure to a pronunciation lexicon FST (that maps sequences of phonemes to sequences of words)~\cite{mohri2002weighted}. However, note that the similarity is only structural; we map graphemes to words, so lexicons are not required. The input vocabulary of $L$ comprises graphemes from the
original alphabet and the output vocabulary of $L$ corresponds to words from the training vocabulary of the ASR system. %For each word in the training data, there is a path in $L$ that, given this word's grapheme sequence as input, produces the word itself as output. 
We also include an \texttt{unk} word in the output vocabulary to accommodate out-of-vocabulary words that might appear in the evaluation data. Any grapheme sequence can map to the \texttt{unk} word, but this is associated with a very large penalty $\eta$. This prevents in-vocabulary words from opting for paths involving \texttt{unk}. 
%Thus, $L$ maps a space-separated grapheme sequence into a sequence of words in the output vocabulary.
\vspace{0.3em}

\noindent \textbf{$G$: Language model FST.} Finally, we have an $N$-gram language model acceptor that rescores word sequences from $H \circ S \circ E \circ L$. The $N$-gram language model is trained on the training set transcriptions of the full speech data.%
\footnote{We observed no additional benefits in performance with using larger text corpora to train $G$.}
\section{Experiments and Results}
%We evaluate the efficacy of our proposed \rnr approach for 2 Indian languages, Gujarati and Telugu using 2 training set durations (Full and $10$ hrs).
\paragraph*{Dataset Details.} We use the Microsoft Speech Corpus (Indian Languages) dataset~\cite{microsoftspeech}~\cite{asrindian2018}. The Gujarati and Telugu speech corpora comprise 39.1 hours and 31.3 hours of training speech, respectively. The dev/test splits for Gujarati and Telugu contain 500/3075 and 500/3040 utterances, respectively. The OOV rates for Gujarati and Telugu on the test set are $5.2\%$ and $12.0\%$, respectively. 
%
%
%
%We use open-source Wikipedia datasets({\Large CITE}) consisting of text-only data for the recurrent neural network-based reconstruction experiments (for details, see \href{tab:text-data}{Table num}). This is also referred to as the \textit{large} dataset in our experiments. We use pretrained FastText embeddings \cite{grave2018learning} for initializing the target language embedding layer for the neural network-based reconstruction experiments when using the grapheme-word tokenization (for details, see \href{tab:embedding-data}{Table num}).

%We construct train, dev, and test sets from the small and large datasets using a foo:bar:raz split. Table num({\Large Ask about this table}) shows detailed statistics for each split.
%Word-OOV rates for dev and test sets can be seen in \href{tab:oov-rates}{Table num}. Vocabulary sizes of the training set post-reduction for various reduction schemes can be seen in \href{tab:vocab-sizes-reduction}{Table num}.

\paragraph*{Experimental Setup.} ASR systems for both languages are built using the ESPNet Toolkit~\cite{watanabe2018espnet}. We use $80$-dimensional log-mel acoustic features with pitch information. Our ASR model is a hybrid LSTM-based CTC-attention model~\cite{watanabe2017hybrid}. %that uses an `identity' reduction and uses the training transcriptions in the original alphabet. 
All our ASR systems use BPE tokenization~\cite{Sennrich_2016}, with BPE-based output vocabularies of size $5000$. We use a CTC weight of $0.8$ and an attention weight $0.2$ with a dropout rate of $0.2$. For both languages, we use 4 encoder layers, 1 decoder layer and location-based attention with $10$ convolutional channels and $100$ filters. For Gujarati, we use $512$ encoder units, $300$ decoder units and an attention dimension of $320$. For Telugu, we use $768$ encoder units, $450$ decoder units and an attention dimension of $250$. We use a beam decoder during inference with a beam width of 45. All models are trained on an Nvidia GeForce GTX 1080 Ti GPU.

All the reconstruction FSTs are implemented using the OpenFST toolkit~\cite{10.5555/1775283.1775287}. The $G$ FST represents a Kneser-Ney smoothed 4-gram LM that is trained using the SRILM toolkit~\cite{stolcke2002srilm}. An edit distance of $d=3$ and a cost of $\lambda=5$ were the best values obtained for the WERs using the full training set duration, by tuning on the dev set. 

% The encoder-decoder reconstruction models were implemented using the Fairseq toolkit~\cite{ott2019fairseq}. For the \gtow tokenization, we use a total vocabulary of $120K$ words in Gujarati and $150K$ words in Telugu. The word embeddings were initialized with pretrained FastText embeddings~\cite{grave2018learning}. For the \btob tokenization, we use a BPE vocabulary of $25000$ for both languages. We used $300$-dimensional embedding layers at the input and output. We used $3$, $512$-unit encoder layers and $3$, $256$-unit decoder layers. We used a dropout rate of $0.2$. During inference, we used a beam decoder with a beam width of $30$.

\subsection{ASR Experiments}
\label{sec:asrexpts}
Table~\ref{tab:asrresults} lists reduced WERs for both Gujarati and Telugu on the dev and test sets using different reduction functions and trained on two different train durations; ``Full" refers to the complete train set and ``10 hr" refers to a randomly sampled 10 hour subset. Recall that for a given reduction function, we train an ASR system with the reduction applied to the ground truth text. Since the above WERs are computed between the hypotheses and the \textit{reduced} ground truth text, and because the output alphabet is different for each reduction function, these are not standard WERs but are rather \textit{reduced} WERs. %since they are computed with respect to a reduced ground truth. 
%and a 10-hr training subset). The 10-hr subset is used to simulate an ultra-low-resource setting.
\begin{table}[t!]
  \caption{Reduced WERs (r-WER) on Gujarati and Telugu for different training set durations.}
  \label{tab:asrresults}
  \centering
  \tabcolsep=0.11cm
  \begin{tabular}{ c | c | c | c | c | c }
    \toprule
    \multicolumn{1}{c}{\textbf{Duration}} &
    \multicolumn{1}{c}{\textbf{Reduction}} &
    \multicolumn{2}{c}{\textbf{r-WER (Guj)} } & 
    \multicolumn{2}{c}{\textbf{r-WER (Tel)} } \\
    \midrule
    & & \textbf{Dev} & \textbf{Test} & \textbf{Dev} & \textbf{Test} \\
    \multirow{3}{*}{Full} & identity & $41.5$ & $43.2$ & $44.1$ & $46.8$ \\
    & \rone & $36.5$ & $39.6$ & $39.3$ & $42.8$ \\
    & \rone-{rand} & $41.3$ & $42.3$ & $44.2$ & $47.9$ \\
    \midrule
     \multirow{3}{*}{10 hr} & identity & $60.2$ & $68.6$ & $64.1$ & $71.4$ \\
    & \rone & $53.9$ & $63.6$ & $56.9$ & $66.5$ \\
    & \rone-{rand} & $63.2$ & $71.8$ & $60.8$ & $69.4$ \\
    \bottomrule
  \end{tabular}
\end{table}

Identity refers to the baseline system with an unaltered grapheme set. \rone is the reduction function discussed in Section~\ref{sec:ourapproach}. \rone-rand is designed to show the importance of a linguistically meaningful reduction. In \rone-rand, graphemes are randomly merged together while ensuring that the size of the final alphabet matches the size of the alphabet after applying the reduction \rone. 

The baseline identity WERs are comparable to previously published results~\cite{Shetty2020ImprovingTP} using LSTM-based E2E architectures for the Microsoft Speech Corpus (Indian Languages) dataset. 
% It is important to note here that the WERs computed for the identity system are not directly comparable to the WERs computed using the other two reduction methods, since they are computed on different vocabularies (original vs. \redalph).
The r-WER for \rone is lower than the r-WER from the identity system, which shows that the ASR system performs an easier task while training on the reduced alphabet. However, the r-WER of \rone-rand is significantly worse than \rone, confirming our claim that linguistically motivated reductions are key to derive performance improvements. 

\begin{table*}[h]
  \caption{WERs from the FST-based reconstruction model for Gujarati and Telugu for two training durations.}
  \label{tab:fst}
  \centering
  \begin{subtable}[t]{0.4\textwidth}
%   \centering
    \tabcolsep=0.11cm
  \begin{tabular}{ c | c | c | c c | c c }
    \toprule
    \multicolumn{1}{c}{$\mathbf{d}$} &
    \multicolumn{1}{c}{$\mathbf{\lambda}$} &
    \multicolumn{1}{c}{\textbf{Reduction}} &
    \multicolumn{2}{c}{\textbf{WER (Guj) }} & 
    \multicolumn{2}{c}{\textbf{WER (Tel) }}\\
    
    \midrule
    
    \multicolumn{3}{c}{} & \textbf{Dev} & \textbf{Test} & \textbf{Dev} & \textbf{Test}\\
   \multicolumn{3}{c}{Baseline} & $41.5$ & $43.2$ & $44.1$ & $46.8$ \\
   
    \cmidrule{1-7}
    
    \multirow{2}{*}{$0$}
    & \multirow{2}{*}{$5$}
    & identity & $41.8$ & $43.4$ & $45.1$ & $47.7$ \\
    & & \rone & $40.4$ & $41.9$ & $42.1$ & $45.7$ \\
    
    % \cmidrule{1-7}
    
    % \multirow{2}{*}{$1$}
    % & \multirow{2}{*}{$5$}
    % & identity & $40.0$ & $40.7$ & $43.3$ & $45.5$ \\
    % & & \rone & $39.2$ & $39.0$ & $40.8$ & $43.7$ \\
    
    % \cmidrule{1-7}
    
    % \multirow{6}{*}{$2$}
    % & \multirow{2}{*}{$2.5$}
    % & identity & $42.7$ & $40.3$ & $43.0$ & $44.7$ \\
    % & & \rone & $43.7$ & $39.7$ & $42.4$ & $44.2$ \\
    
    % &
    % \multirow{2}{*}{$5$}
    % & identity & $38.4$ & $38.7$ & $41.3$ & $43.4$ \\
    % & & \rone & $37.9$ & $37.1$ & $39.1$ & $41.9$ \\
    
    % &
    % \multirow{2}{*}{$10$}
    % & identity & $39.2$ & $40.1$ & $41.9$ & $44.1$ \\
    % & & \rone & $38.5$ & $39.1$ & $39.6$ & $42.4$ \\
    \cmidrule{1-7}
    
    \multirow{2}{*}{$3$}
    & \multirow{2}{*}{$5$}
    & identity & $37.9$ & $37.8$ & $40.6$ & $42.5$ \\
    & & \rone & $\textbf{37.8}$ & $\textbf{36.5}$ & $\textbf{38.5}$ & $\textbf{41.2}$ \\
    \bottomrule
  \end{tabular}
  \caption{Full training duration.}
  \end{subtable}
  \quad
  \begin{subtable}[t]{0.4\textwidth}
  \tabcolsep=0.11cm
  \begin{tabular}{ c | c | c | c c | c c }
    \toprule
    \multicolumn{1}{c}{$\mathbf{d}$} &
    \multicolumn{1}{c}{$\mathbf{\lambda}$} &
    \multicolumn{1}{c}{\textbf{Reduction}} &
    \multicolumn{2}{c}{\textbf{WER (Guj) }} & 
    \multicolumn{2}{c}{\textbf{WER (Tel) }}\\
    
    \midrule
    
    \multicolumn{3}{c}{} & \textbf{Dev} & \textbf{Test} & \textbf{Dev} & \textbf{Test}\\
   \multicolumn{3}{c}{Baseline} & $60.2$ & $68.6$ & $64.1$ & $71.4$ \\
   
    \cmidrule{1-7}
    
    \multirow{2}{*}{$0$}
    & \multirow{2}{*}{$5$}
    & identity & $60.3$ & $68.6$ & $64.4$ & $71.6$ \\
    & & \rone & $56.2$ & $64.9$ & $58.4$ & $67.8$ \\

    \cmidrule{1-7}
    
    \multirow{2}{*}{$3$}
    & \multirow{2}{*}{$5$}
    & identity & $56.8$ & $64.9$ & $59.2$ & $66.1$ \\
    & & \rone & $\textbf{53.2}$ & $\textbf{61.2}$ & $\textbf{54.3}$ & $\textbf{63.6}$ \\
    \bottomrule
  \end{tabular}
\caption{10-hr training duration.}
  \end{subtable}
\end{table*}

\subsection{FST-based Reconstruction Experiments}
\label{sec:fstrecon}
Table~\ref{tab:fst} shows the WERs for both Gujarati and Telugu using our FST-based reconstruction model with the two training durations. Recall that $d$ is the edit distance permitted by the edit distance FST and $\lambda$ is the associated edit cost. $d=0$ thus means only allowing for words in the original alphabet that can be exactly recovered from the reconstructed word forms. We find that in this scenario, reconstruction with the \rone mapping for $d=0$ significantly outperforms both the baseline and the identity reduction. As we increase $d$ from $0$ to $3$, the power of reconstruction increases due to the $E$ FST and we observe large reductions in WER for both \rone and identity. Even with this best reconstruction system, the \rone mapping has significantly lower WERs than the corresponding identity mapping. In the $d=3, \lambda=5$ setting, averaged across both languages, we observe a $3.2\%$ relative test WER decrease for the Full duration and a $4.8\%$ relative test WER decrease for the 10-hr duration. The benefits of our approach are more pronounced in the 10 hour setting, reaffirming its utility for low-resource languages.
\subsection{Reconstruction after RNNLM rescoring}

%RNNLM rescoring during decoding is a common technique to improve the quality of E2E ASR systems. 
We examine whether \rnr is effective even after using an RNNLM rescoring during decoding. We train a 2-layer RNNLM (with $1500$ hidden units each) on the training transcriptions of the full speech data to rescore the predictions from the ASR systems during beam decoding. Table~\ref{tab:rnnlm} lists WERs for both Gujarati and Telugu using the best reconstruction system ($d=3, \lambda=5$) for both training durations. With the RNNLM in place, we observe a significant improvement in performance of the baseline system that uses no reconstruction. In the 10-hr setting, our approach using $\rho_1$ performs significantly better than the baseline and the identity mapping for both languages, yielding up to $7\%$ relative reduction in test WERs. In the full setting, Gujarati does not benefit from \rnr while Telugu still yields improvements on test WER. %Here, one observes a meager $0.7\%$ relative test WER decrease (averaged across languages). 
It is clear that the improvements owing to \rnr are more pronounced in very low-resource settings. This trend was observed in Table~\ref{tab:fst} as well. 
\begin{table}[H]
  \caption{WERs for Gujarati and Telugu using $d=3, \lambda=5$, in conjunction with RNNLM rescoring}
  \label{tab:rnnlm}
  \centering
  \tabcolsep=0.11cm
  \begin{tabular}{c | c | c | c | c | c }
    \toprule
    \multicolumn{1}{c}{\textbf{Duration}} &
    \multicolumn{1}{c}{\textbf{Reduction}} &
    \multicolumn{2}{c}{\textbf{WER (Guj)} } & 
    \multicolumn{2}{c}{\textbf{WER (Tel)} } \\
    \midrule
    & & \textbf{Dev} & \textbf{Test} & \textbf{Dev} & \textbf{Test} \\
    \multirow{3}{*}{Full}
    & Baseline & $37.4$ & $34.0$ & $37.9$ & $40.0$ \\
    \cmidrule{2-6}
    & identity  & $\textbf{36.2}$ & $\textbf{31.8}$ & $37.7$ & $39.2$ \\ 
    & \rone & $37.1$ & $32.2$ & $\textbf{36.5}$ & $\textbf{38.1}$ \\
    %\multirow{2}{*}{CTC+Attention+RNNLM}
    %& identity & $37.4$ & $34.0$ \\
    %& reduced-1 & $34.4$ & $32.7$ \\
    \midrule
    \multirow{3}{*}{10-hr}
    & Baseline & $56.2$ & $63.2$ & $56.9$ & $63.8$ \\
    \cmidrule{2-6}
    & identity  & $55.5$ & $62.3$ & $56.2$ & $62.5$ \\ 
    & \rone & $\textbf{52.0}$ & $\textbf{58.2}$ & $\textbf{51.2}$ & $\textbf{59.1}$ \\
    %\multirow{2}{*}{CTC+Attention+RNNLM}
    %& identity & $37.4$ & $34.0$ \\
    %& reduced-1 & $34.4$ & $32.7$ \\
    \bottomrule
  \end{tabular}
\end{table}
%Thus, as the amount of training data available to the system decreases i.e. as the language is increasingly low-resource, the improvements due to the \rnr approach become more significant. This trend was observed in the non-RNNLM setting in Section~\ref{sec:fstrecon} as well. This shows that our approach is well suited to low-resource settings

% We further examine why Telugu is benefiting from \rone  despite having higher OOV rates. In the Telugu vocabulary, roughly $35K$ reduced word forms deterministically map to a single word in the original alphabet, while this number is less than $30K$ for Gujarati. Thus, there is less ambiguity in Telugu reconstruction overall. Also, all the plosives together make up $35.5\%$ and $33.4\%$ of the total number of graphemes in Gujarati and Telugu, respectively. Smaller number of grapheme merges in Telugu could be another reason for Telugu's improved performance with \rnr.

%\todo[inline]{Should we remove the justification of why Guj didn't work, since we don't need to justify it now? Commented it out for now.}

\begin{table}[H]
  \caption{Results for Gujarati 10-hr using a conformer model}
  \label{tab:asrresultsconformer}
  \centering
  \tabcolsep=0.05cm
  \begin{subtable}[t]{0.2\textwidth}
  \begin{tabular}{c | c | c}
    \toprule
    \multicolumn{1}{c}{\textbf{Reduction}} &
    \multicolumn{2}{c}{\textbf{r-WER (Guj)} }\\
    \midrule
    & \textbf{Dev} & \textbf{Test} \\
    identity & $57.7$ & $61.1$ \\
    \rone & $56.4$ & $59.5$ \\
    \rone-{rand} & $59.6$ & $62.4$ \\
    \bottomrule
  \end{tabular}
   \caption{r-WER on ASR predictions}
  \end{subtable}
  \quad
  \begin{subtable}[t]{0.2\textwidth}
   \begin{tabular}{ c | c | c | c c }
    \toprule
    \multicolumn{1}{c}{$\mathbf{d}$} &
    \multicolumn{1}{c}{$\mathbf{\lambda}$} &
    \multicolumn{1}{c}{\textbf{Reduction}} &
    \multicolumn{2}{c}{\textbf{WER (Guj) }} \\
    
    \midrule
    
    \multicolumn{3}{c}{} & \textbf{Dev} & \textbf{Test} \\
   \multicolumn{3}{c}{Baseline} & $57.7$ & $61.1$ \\
   
    \cmidrule{1-5}
    
    \multirow{2}{*}{$0$}
    & \multirow{2}{*}{$10$}
    & identity & $57.7$ & $61.1$ \\
    & & \rone & $57.9$ & $60.4$ \\

    \cmidrule{1-5}
    
    \multirow{2}{*}{$3$}
    & \multirow{2}{*}{$10$}
    & identity & $\textbf{57.1}$ & $60.5$ \\
    & & \rone & $57.6$ & $\textbf{59.9}$ \\
    \bottomrule
  \end{tabular}
  \caption{WERs after reconstruction}
  \end{subtable}
\end{table}
\subsection{Using conformers}
To demonstrate that \rnr is effective across different E2E architectures, we also train a conformer-based model~\cite{gulati2020conformer} for Gujarati 10-hours shown in Table~\ref{tab:asrresultsconformer}. 
These models were also built using the ESPNet Toolkit~\cite{watanabe2018espnet}. We use 2 encoder layers and 1 decoder layer, each with 350 units and 4 attention heads. The encoder output has 256 units. We use a CTC weight of 0.3 with a dropout rate of 0.1. Following similar trends as in the hybrid models, we find \rone to be much more effective as a reduction compared to \rone-rand and reconstruction with \rone yields WER improvements over the identity system on the test set. 

\section{Discussion and Qualitative Analysis}
\subsection{Choice of reduction function}
%The reconstruction module larger text did not benefit the FST. 
%\noindent \textbf{Larger text corpus.}
One might wonder about the impact of the reduction function on ASR performance. In Table~\ref{tab:asrresults}, we show how a randomly chosen \rone-rand can be detrimental to the ASR system. Additionally, we explore employing a third reduction function \rtwo that is less compressive than \rone. \rtwo reduces pairs of aspirated/unaspirated plosives into a single grapheme rather than reducing quadruplets of plosives (as in \rone) and does not merge long and short vowel variants, resulting in a \redalph of size $48$ for Gujarati and size $55$ for Telugu. With \rtwo, on the 10-hr training set with a $(d=0, \lambda=5)$ system, we obtain WERs of $67.8\%$ and $67.9\%$ on the Gujarati and Telugu test sets, respectively, which is worse compared to $64.9\%$ and $67.8\%$ obtained with \rone. This shows that a reduction function like \rone that results in reduced alphabets with no underlying acoustically confusable sounds is more beneficial with \rnr than other reduction functions.

\subsection{Effect of reduction function on correcting ASR errors}
Consider the pre-reconstruction ASR system (from Section~\ref{sec:asrexpts}) trained on the 10-hr training subset. %We obtain the identity and reduced predictions for the test set (used to compute the r-WERs in Table~\ref{tab:asrresults}) and individually compute character alignments of each prediction with the reference transcriptions to analyze the nature of substitution errors made by the identity ASR system versus the reduced ASR system.
Of the substitution errors made by the identity system, we compute the percentage of errors that were corrected by the reduced predictions. To do this, we first align the identity hypothesis text to the reference text. Call this alignment $a_1$. We similarly align the reduced hypothesis text to the reduced reference text. Call this alignment $a_2$. Let $X$ be the number of substitution errors made by the identity system i.e., the number of character substitutions $a \shortrightarrow b$ (where $a \neq b$) in the identity system alignment $a_1$. Out of these substitutions, we count the number of substitutions $Y$ for which the corresponding character alignment in $a_2$ is correctly predicted. This percentage is calculated as $\frac{Y}{X} \times 100 \%$. Of all the substitution errors made by the identity system, $16.29 \%$ (for Gujarati) and $16.92 \%$ (for Telugu) of the errors were corrected by the reduced predictions. This shows that the reduced system is able to fix a sizable number of substitution errors incurred by the identity system.

\subsection{Test set perplexities before and after reduction}
As an information-theoretic measure of reduction , we compute the perplexity (ppl) of the test set using a trigram LM trained on the training set, in both the original and reduced vocabularies in Table~\ref{tab:ppl}. The ppl reductions show that our reduced vocabulary makes the language model-based predictions more accurate; the larger drop in ppl for Telugu also correlates with the larger improvement in WERs for Telugu compared to Gujarati.
\begin{table}[H]
  \caption{Test set perplexities using a trigram LM}
  \label{tab:ppl}
  \centering
  \tabcolsep=0.11cm
  \begin{tabular}{c | c | c }
    \toprule
    \textbf{Reduction} & \textbf{Test ppl (Guj)} & \textbf{Test ppl (Tel)} \\
    \midrule
    identity & $115.05$ & $768.66$ \\
    \rone & $108.13$ & $706.32$ \\
    \bottomrule
  \end{tabular}
\end{table}
%
% \begin{figure}[b!]
% \caption{An illustrative example each from Gujarati and Telugu using FST reconstruction, $\mathbf{d}=3,\mathbf{\lambda}=5$. R: Reference, I: Identity reduction, \rone: Our reduction}
%     \begin{subfigure}{0.25\textwidth}
%     \includegraphics[width=0.5\linewidth]{guj.png}
%     \end{subfigure}
%     \begin{subfigure}{0.25\texwidth}
%     \includegraphics[width=0.5\linewidth]{tel.png}
%     \end{subfigure}
% \label{fig:examples}
% \end{figure}
%

\subsection{Illustrative Examples}
\begin{figure}[h!]
    \caption{An illustrative example each from Gujarati and Telugu using FST reconstruction, $\mathbf{d}=3,\mathbf{\lambda}=5$. $R$ and $I$ are the reference and identity transcriptions.}
    \centering
    \includegraphics[width=0.99\linewidth]{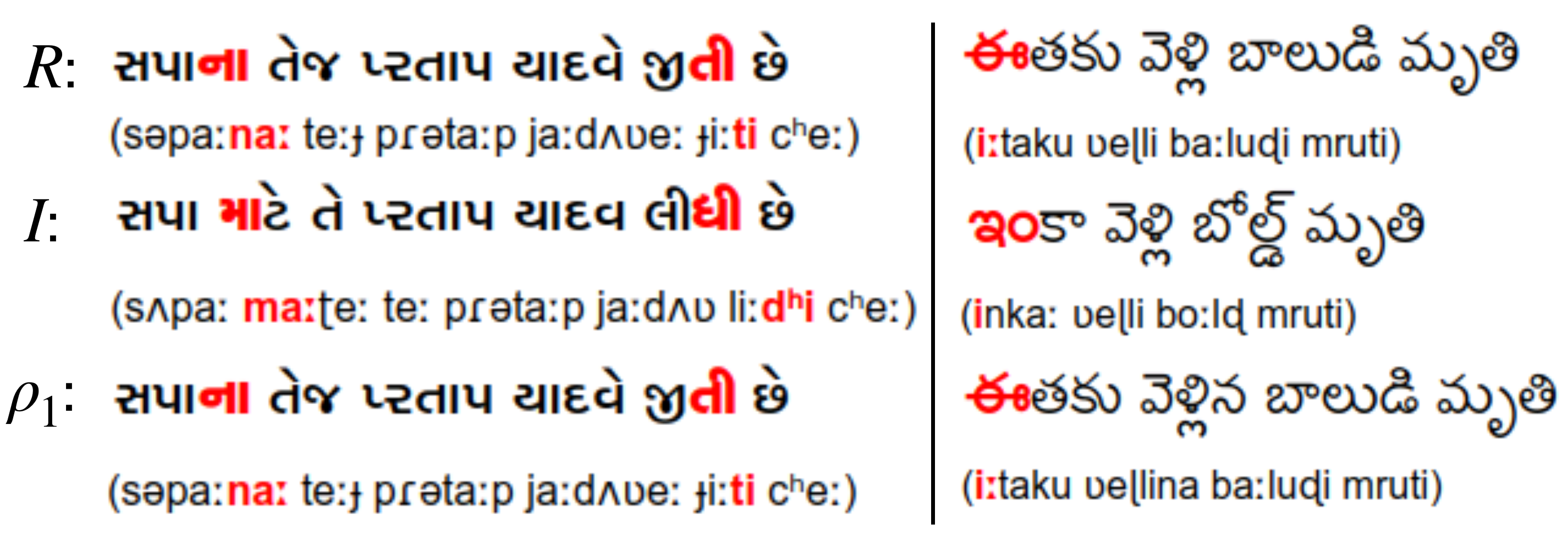}
    \label{fig:examples}
\end{figure}

Figure~\ref{fig:examples} shows two examples of \rnr. The first example highlights nasal/plosive sounds in Gujarati that were correctly recognized by \rone, while the identity system recognizes a different nasal/plosive sound that is acoustically confusable and part of the reduction map in \rone. The second example shows a long vowel in Telugu mistakenly recognized as a short vowel; here again, \rone merges these two graphemes and hence accurately recovers the correct form.
%*}

\section{Conclusions}
This work proposes an effective reduce-and-reconstruct technique that can be used with any ASR system for low-resource phonetic languages. We demonstrate its utility for two Indian languages and show that as the available training data decreases, our approach yields greater benefits, making it well-suited for low-resource languages. Future work will investigate the use of more powerful Transformer-based~\cite{NIPS2017_3f5ee243} reconstruction models and automatically learning a reduction mapping based on errors made by the ASR system.

\newpage
\bibliography{main}
\bibliographystyle{IEEEtran}
% \input{main.bbl}
% \fi

% \begin{thebibliography}{9}
% \bibitem[1]{Davis80-COP}
%   S.\ B.\ Davis and P.\ Mermelstein,
%   ``Comparison of parametric representation for monosyllabic word recognition in continuously spoken sentences,''
%   \textit{IEEE Transactions on Acoustics, Speech and Signal Processing}, vol.~28, no.~4, pp.~357--366, 1980.
% \bibitem[2]{Rabiner89-ATO}
%   L.\ R.\ Rabiner,
%   ``A tutorial on hidden Markov models and selected applications in speech recognition,''
%   \textit{Proceedings of the IEEE}, vol.~77, no.~2, pp.~257-286, 1989.
% \bibitem[3]{Hastie09-TEO}
%   T.\ Hastie, R.\ Tibshirani, and J.\ Friedman,
%   \textit{The Elements of Statistical Learning -- Data Mining, Inference, and Prediction}.
%   New York: Springer, 2009.
% \bibitem[4]{YourName17-XXX}
%   F.\ Lastname1, F.\ Lastname2, and F.\ Lastname3,
%   ``Title of your INTERSPEECH 2020 publication,''
%   in \textit{Interspeech 2020 -- 20\textsuperscript{th} Annual Conference of the International Speech Communication Association, September 15-19, Graz, Austria, Proceedings, Proceedings}, 2020, pp.~100--104.
% \end{thebibliography}

\end{document}